\documentclass[superscriptaddress,english,aps,pra,twocolumn,groupedaddress,floatfix]{revtex4}

\usepackage{amsmath}
\usepackage{amssymb}
\usepackage{amsfonts}
\usepackage{bbm}
\usepackage{epsfig}
\usepackage{times}
\usepackage{babel}

\usepackage{color}

\newcommand{\Tr}{\operatorname{Tr}}
\newcommand{\bra}{\langle}
\newcommand{\ket}{\rangle}
\renewcommand{\vec}[1]{\boldsymbol{#1}}
\newcommand{\mc}[1]{\mathcal{#1}}

\newcommand{\phys}{{\text{phys}}}

\newcommand{\peps}{{\text{PEPS}}}
\newcommand{\refined}{{\text{refined}}}
\newcommand{\gs}{{\text{gs}}}
\newcommand{\mL}{\mc{L}}
\newcommand{\V}{\mc{V}}
\newcommand{\E}{\mc{E}}
\newcommand{\A}{\mc{A}}

\newcommand{\potsdam}{Institute for Physics and Astronomy, Potsdam University, 14476 Potsdam, Germany}
\newcommand{\wiko}{Institute for Advanced Study Berlin, 14193 Berlin, Germany}

\begin{document}

\title{Real-space renormalization yields finite correlations}

\author{Thomas Barthel}
\affiliation{\potsdam}
\author{Martin Kliesch}
\affiliation{\potsdam}
\author{Jens Eisert}
\affiliation{\potsdam}
\affiliation{\wiko}
\date{March 01, 2010}
 
\begin{abstract}
Real-space renormalization approaches for quantum lattice systems generate certain hierarchical classes of states that are subsumed by the multi-scale entanglement renormalization ansatz (MERA). It is shown that, with the exception of one spatial dimension, MERA states are actually finitely correlated states, i.e., projected entangled pair states (PEPS) with a bond dimension independent of the system size. Hence, real-space renormalization generates states which can be encoded with local effective degrees of freedom, and MERA states form an efficiently contractible class of PEPS that obey the area law for the entanglement entropy. It is shown further that there exist other efficiently contractible schemes violating the area law.
\end{abstract}

\maketitle

\section{Introduction}
\begin{figure}
\centering
\includegraphics[width=1\linewidth]{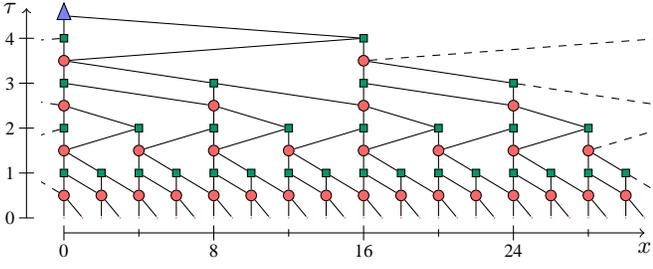}
\caption{\label{fig:MERA1d}
A 1D MERA with linear branching ratio $b=2$. Circles, squares, and the triangle denote tensors, the lines denote contractions of those tensors. The squares are (partial) isometries that map two local subsystems $\mc H_{i}^{\tau}$ and $\mc H_{i+1}^{\tau}$ into one $\mc H_{i/2}^{\tau+1}$ as in Kadanoff's block spin transformation. The circles denote unitary operators, \emph{disentanglers}, that reduce the entanglement between $\mc H_{i}^{\tau}\otimes \mc H_{i+1}^{\tau}$ and the rest of the system before the action of the isometry.
When, for the mapping to a 1D PEPS, MERA tensors are assigned to lattice sites according to Eq.~\eqref{eq:tensorPos-Wrong}, as in the diagram, \emph{stacks} of tensors occur: There are sites where the number of assigned MERA tensors diverges with the system size, implying that the mapping is inefficient. The stacking of tensors can be avoided by choosing tensor positions according to Eq.~\eqref{eq:tensorPos}.}
\end{figure}

Renormalization group (RG) methods aim at solving many-body problems by treating energy scales in an iterative fashion, progressing from high to low energies \cite{Wilson1975,Wegner1972-5}. One of its earliest formulations is the real-space RG which works by repeated steps of thinning out local degrees of freedom and rescaling of the system as in Kadanoff's block spin transformation \cite{Kadanoff1966-2}.
In real-space RG approaches to quantum lattice models \cite{Jullien1977-38,Drell1977-16}, in each RG step $\tau$, the system is partitioned into small blocks. From those blocks high-energy states are eliminated and the Hamiltonian $\hat H_{\tau+1}$ for the renormalized system is obtained by applying the corresponding projection operators, exactly $\hat H_{\tau+1}=\hat P_{\tau+1} \hat H_\tau \hat P_{\tau+1}^\dag$ or in some appropriate approximation, followed by a coarse-graining of the lattice. This is iterated, e.g., until a step $\tau=T$ is reached where the renormalized system consists of a single small block for which the ground-state $|\gs_T\ket$ can be obtained exactly. Applying the RG transformations in reverse order yields an approximation $\hat P_1^\dag \hat P_2^\dag\dotsm \hat P_T^\dag|\gs_T\ket$ to the ground-state of the original model. Those states, generated by the real-space RG, fall into the class of so-called \emph{tree tensor networks} (TTN) \cite{Shi2006-74}.
A recent more elaborate real-space RG scheme, the \emph{multi-scale entanglement renormalization ansatz} (MERA) \cite{Vidal-2005-12,Vidal2006}, a genuine simulation technique for strongly correlated systems, allows in each RG step for local unitary operations to be applied before the elimination of block basis states
\footnote{Another real-space RG scheme for the simulation of strongly correlated systems is the \emph{contractor renormalization group} (CORE) \cite{Morningstar1994-73,Morningstar1996-54}. The underlying idea is similar to MERA in the sense that it tries to take inter-block correlations better into account than the earlier RG schemes \cite{Jullien1977-38,Drell1977-16} and that, in each RG step, $k$-local Hamiltonians are mapped to $k$-local Hamiltonians by application of local isometries. However, CORE focuses entirely on the flow of Hamiltonians. As it does not generate a (unique) ground-state approximation, the method is not a topic of this article.}.
The technique generates hence a more general class of (variational) states, referred to as MERA states; see Fig.~\ref{fig:MERA1d}.

Whereas the degrees of freedom of MERA and TTN states are organized in a hierarchical structure encoding correlations on different length scales, there exists a different class of so called \emph{finitely correlated states} where the degrees of freedom are organized in a strictly local manner. For $D=1$ dimensional systems they are often referred to as \emph{matrix product states} \cite{Accardi1981,Fannes1992-144,Rommer1997}, and for $D\geq 1$ as \emph{tensor product ans\"atze} or \emph{projected entangled pair states} (PEPS) \cite{Niggemann1997-104,Nishino2000-575,Martin-Delgado2001-64,Verstraete2004-7,Verstraete2006-96}; see Fig.~\ref{fig:TCS-to-PEPS}. PEPS are the basis of powerful numerical techniques, such as the extraordinarily successful \emph{density-matrix renormalization group} method \cite{White1992-11,Schollwoeck2005}.

In this work, we establish the surprising fact that, for \mbox{$D>1$}, real-space RG, despite of the inherently hierarchical nature of the procedure, generates states that capture correlations by \emph{local} degrees of freedom. More specifically, it is shown that MERA states form a subclass of PEPS, unifying both approaches. This also explains the failure of real-space RG for some situations for which merely anecdotal evidence had previously accumulated.

PEPS, TTN, and MERA are all \emph{tensor network states} (TNS). In terms of an orthonormal product basis $|\vec{\sigma}\ket=\bigotimes_{i=1}^N|\sigma_i\ket$ for a lattice of $N$ sites, TNS are of the form $\sum_{\vec{\sigma}} \psi_{\vec{\sigma}}|\vec{\sigma}\ket$ where the expansion coefficients $\psi_{\vec{\sigma}}$ are encoded as a partially contracted set of tensors; Fig.~\ref{fig:TCS-to-PEPS}. Recently, this notion has been generalized to the fermionic case \cite{Kraus2009_04,Corboz2009_04,Pineda2009_05,Barthel2009-80}.
For a PEPS, to each site $i$, a tensor $A_i$ is assigned which has one \emph{physical index} $\sigma_i$ and further \emph{auxiliary indices} -- one for each nearest neighbor -- which need to be contracted to obtain $\psi_{\vec{\sigma}}$; Fig.~\ref{fig:TCS-to-PEPS}. For TTN and MERA, the tensors are arranged in a hierarchical pattern with the physical indices in the lowest layer; Fig.~\ref{fig:MERA1d}.
The number of degrees of freedom of a TNS can be tuned by changing the number $\chi$ of values each auxiliary index runs over. Increasing $\chi$ for a fixed structure of the TNS, enlarges the variational space, allowing for a more precise approximation to the exact ground-state in a variational method, but increases computation costs. Hence, $\chi$ is called the \emph{refinement parameter} of the TNS. The computational costs for efficient simulation techniques scale polynomially in $\chi$.

In this article the following is shown.
\begin{itemize}
	\item For $D>1$ spatial dimensions, MERA states with a refinement parameter $\chi$ can be mapped efficiently to PEPS such that the resulting PEPS refinement parameter $\chi_\peps$ is a system-size  independent function of $\chi$.
	\item For $D=1$, TTN and MERA states can in general not be encoded efficiently as PEPS. For a fixed $\chi$ there are MERA states with an entanglement entropy that scales logarithmically in the system size, as occurring in critical models \cite{Amico2008-80,Eisert2008}. In this sense MERA states are more useful than PEPS for this case.
	\item For $D>1$, there exist other efficiently contractible TNS, based on \emph{quantum cellular automata}, that generate an amount of entanglement violating the area law.
\end{itemize}
The ability to map MERA states efficiently to PEPS for $D>1$ implies that $D>1$ MERA states always obey the entanglement \emph{area law} just as PEPS \cite{Amico2008-80,Eisert2008,Vidal2006arXiv}. This behavior is shared by ground-states of non-critical systems and critical bosons. Ground-states of critical fermions, however, can violate the area law \cite{Wolf2005,Gioev2005,Barthel2006-74,Li2006,Cramer2006,Amico2008-80,Eisert2008}. Consequently $\chi$ needs to be scaled polynomially in the system size in order to describe such critical fermionic systems accurately. Otherwise, the real-space RG schemes addressed here \cite{Jullien1977-38,Drell1977-16,Vidal-2005-12} are imprecise in that case.
The remaining advantage of $D>1$ MERA is that local observables and correlation functions can be evaluated efficiently, whereas, for PEPS, approximations are necessary. In this sense, MERA states form an efficiently contractible subclass of PEPS. 

\section{General procedure for mapping TNS to PEPS}\label{sec:genTCStoPEPS}
\begin{figure}
\centering
\includegraphics[width=1\linewidth]{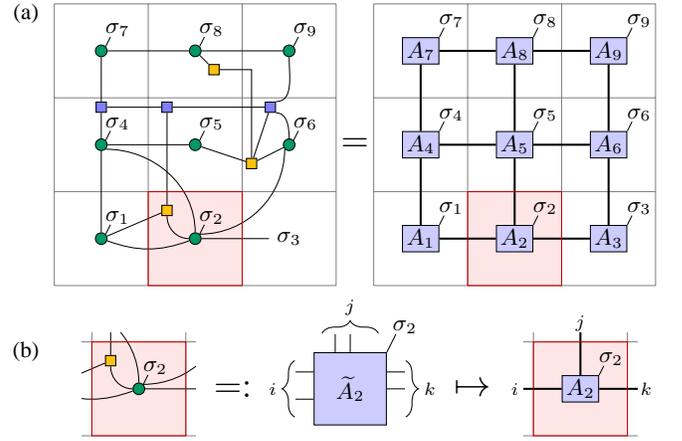}
\caption{\label{fig:TCS-to-PEPS}
(a) Procedure for mapping a TNS (left) to a 2D PEPS (right), by assigning tensors to lattice sites and contraction lines to paths on the lattice.
(b) The elements of the PEPS tensors are determined by the elements of the tensors composing the TNS.}
\end{figure}

All TNS can be mapped to PEPS, although not necessarily in an efficient manner. To map a TNS to a PEPS one can
\begin{enumerate}
	\item assign each tensor of the TNS to a specific site of the physical lattice \footnote{For clarity we restrict to square lattices.}
	\begin{equation}
		\V_\phys:=\{0,\dotsc,L-1\}^D\subset \mathbb{Z}^D,
	\end{equation}
	and,
	\item for each contraction line that connects the tensors, decide on a specific path for that line on the edges $\E_\phys$ of the physical lattice,
	\begin{equation}
		\E_\phys:=\{(\vec{r},\vec{r}')\in\V_\phys\times\V_\phys |\, |\vec{r}-\vec{r}'|_1=1\},
	\end{equation}
\end{enumerate}
see Fig.~\ref{fig:TCS-to-PEPS}.
The tensors composing the PEPS are then obtained by introducing for each edge of the lattice an auxiliary vector space that is the tensor product of the vector spaces of all TNS contraction lines that traverse that edge. The elements of the PEPS tensor for site $i$ are determined by the elements of all the TNS tensors that were assigned to site $i$. See Fig.~\ref{fig:TCS-to-PEPS}b. 

Applying this procedure for a 1D MERA state inevitably results in a PEPS refinement parameter $\chi_\peps$ that diverges with the system size. This is not just a feature of the specific procedure. In Sec.~\ref{sec:1DcounterExample}, a family of 1D TTN states is constructed for which any mapping to PEPS necessarily requires $\chi_\peps$ to diverge with the system size.

\section{Efficiently mapping MERA to PEPS for $D>1$}
Given a family of MERA states for different lattice sizes $L$, a mapping of the MERA states to PEPS is called \emph{efficient} if there exists an upper bound $\chi_\peps$ on the resulting PEPS refinement parameter that is independent of $L$.

\subsection{Qualitative argument for the existence of an efficient mapping of $D>1$ MERA to PEPS}
The following argument motivates why an efficient mapping of MERA to PEPS should be possible for $D>1$ but not for $D=1$.
Let us assign to each contraction line of the MERA state a finite cross-section, e.g., equal to $a^{D-1}$ with the lattice spacing $a$. Then one can ask what $D$-dimensional volume $V(\tau)$ the contraction lines of a certain layer $\tau$ connecting to layers with $\tau'\leq\tau$ cover. Those contraction lines of layer $\tau$ have length $\ell(\tau)\propto a b^\tau$, where $b$ is the \emph{linear branching ratio} of the MERA. The number of lattice cells in layer $\tau$ is $b^{(T-\tau)D}$; Fig.~\ref{fig:MERA1d}. Hence, the volume covered by the contraction lines of layer $\tau$ is $V(\tau)\propto a^{D-1}\ell(\tau) b^{(T-\tau) D}\propto b^{DT-(D-1)\tau}$. The density of the MERA contraction lines, or more precisely, a resulting estimate for the average number of contraction line paths traversing a unit cell of the physical lattice ($\tau=0$) is hence 
\begin{eqnarray}\label{eq:lineDensity}
	\log_\chi(\chi_\peps)&\propto& b^{-TD} \sum_{\tau=0}^T V(\tau)
 \propto \sum_{\tau=0}^T b^{-(D-1)\tau}\nonumber\\
 &\propto&
 \begin{cases}
 	T&\text{for } D=1\\
\frac{1}{1-b^{-(D-1)}}&\text{for }D>1,\, T\to\infty.
 \end{cases}
\end{eqnarray}
Note that for an edge that is traversed by $n$ paths, one obtains an upper bound $\chi_\peps =\chi^n$ to the corresponding PEPS refinement parameter, i.e., $n=\log_\chi(\chi_\peps)$. As $T=\log_b L$, 1D MERA with a fixed refinement parameter $\chi$ have according to Eq.~\eqref{eq:lineDensity} the potential to encode states with a logarithmic scaling of the entanglement entropy, as occurring in critical 1D systems. See Sec.~\ref{sec:1DcounterExample} for an example.
For $D>1$, however, Eq.~\eqref{eq:lineDensity} suggests that there is enough space on the physical lattice to assign the MERA contraction lines to paths on the lattice in such a way that, for a fixed $\chi$, the resulting PEPS has a bond dimension $\chi_\peps$ that is independent of the system size. That this is indeed possible is proven constructively in the following.

\subsection{Preconditions for MERA states}\label{sec:MERAconditions}
In order to show that the mapping presented in the following is efficient, it is necessary to exploit the defining properties of MERA states that correspond directly to features of the real-space RG and can be summarized as follows.
\begin{enumerate}
	\item The MERA state is a TNS for a $D$-dimensional square lattice $(\V_\phys,\E_\phys)$ consisting of $L^D$ unit cells with
	\begin{equation}
		L=b^T.
	\end{equation}
	\item The MERA consists of $T$ layers of tensors labeled by $\tau=1,\dotsc,T$. 
	\item There is an upper bound $\chi$ on the dimension of the vector spaces associated to the tensor indices, and an upper bound $C_o$ on the order of each tensor. 
	\item With each layer, we associate a coarse-grained square lattice $\mL_\tau$ of $(L/b^{\tau})^D$ cells of the physical lattice
	\begin{equation}
 		\mL_\tau:=\{0,\dots,L/b^{\tau}-1\}^D\subset \mathbb{Z}^D,
	\end{equation}
	and $\mL_0:= \V_\phys$. Every cell of lattice $\mL_\tau$ contains corresponding $b^{D}$ cells of lattice $\mL_{\tau-1}$ \footnote{It is assumed here that the branching ratio $b$ is integer. For cases where $b$ is noninteger, it is only important that $\lfloor b\rfloor\geq 2$. Even if the original $b$ is below $2$, one obtains the necessary branching ratio $\geq 2$, by considering a sufficient number of original layers as one layer.}.
	\item There exists an assignment of the tensors of layer $\tau$ to cells of the lattice $\mL_\tau$ such that
	\begin{enumerate}
		\item the number of tensors inside a single cell is bounded from above by a constant $C_t$,
		\item the distance of contracted tensors is bounded from above by $C_r$, where the distance of a tensor of layer $\tau$ to a tensor of layer $\tau'\leq \tau$ is defined as the $L_1$ distance of their corresponding cells in $\mL_{\tau}$ \footnote{The coarse-graining maps cells from $\mL_{\tau'}$ to cells in $\mL_{\tau>\tau'}$.}.
	\end{enumerate}
	\item For $|\tau-\tau'|>C_{T}$, there are no contractions between tensors of layer $\tau$ with tensors of layer $\tau'$.
\end{enumerate}
The upper bounds $\chi$, $C_o$, $C_t$, $C_r$, and $C_T$  are required to be independent of the system size $L$. \footnote{Actually, a system-size independent upper bound on the combination occurring in Eq.~\eqref{eq:chiBound} is sufficient.}
The stated conditions guarantee that the MERA states feature a so-called \emph{causal cone} \cite{Vidal2006}. Hence, local observables can be evaluated efficiently if all tensors are chosen isometric. For the mapping described in the following, this is however not necessary. As we require only upper bounds on the MERA refinement parameter, which may hence actually vary from tensor to tensor, the apparent restriction to square lattices is not essential. The conditions stated above are met for all typical MERA structures considered in the literature so far. In Fig.~\ref{fig:MERA2d}a a 2D MERA with $b=2$, $C_o=8$, $C_t=2$, and $C_T=1$ is depicted. Fig.~\ref{fig:MERA2d}b shows a 2D MERA with $b=3$, $C_o=10$, $C_t=4$, and $C_T=1$. One can reach $C_r=2$ for both examples.

\subsection{The efficient mapping}
Let us explain a general scheme for mapping MERA states for $D$-dimensional systems efficiently to PEPS. The preconditions of Sec.~\ref{sec:MERAconditions} are assumed to be given.
A simple procedure to assign the MERA tensors to certain lattice sites is to put the tensors of cell $\vec{n}\in\mL_\tau$ of layer $\tau$ to the site
\begin{equation}\label{eq:tensorPos-Wrong}
	\vec{r}_{\tau}(\vec{n}) = b^\tau\vec{n} \quad \in \V_\phys 
\end{equation}
as in Fig.~\ref{fig:MERA1d}.
The problem with this approach is that one generates \emph{stacks} of tensors at certain lattice sites, i.e., there exist positions $\vec{r}\in \V_\phys$ to which a number of tensors is assigned that is not independent of the lattice size. For example, at site $\vec{r}=(0,\dotsc,0)$ a number of $\propto T = \log_b L$ tensors accumulate.  Further stacks of tensors with height $\propto T'$ accumulate for example at lattice sites with coordinates $b^{T'}(1,\dotsc,1)\in\V_\phys$, see Fig.~\ref{fig:MERA1d}.
It is necessary to avoid such stacks of tensors, because they imply in general that $\chi_\peps$ diverges with the system size. Stacks can be avoided by shifting the allowed tensor positions for different layers relative to each other. One possible such choice for $\vec{r}_{\tau}(\vec{n})\in \V_\phys$ is
\begin{equation}\label{eq:tensorPos}
	\vec{r}_{\tau}(\vec{n}) = b^\tau\vec{n}+b^{\tau-1}\vec{e} \quad\text{with}\quad \vec{n}\in\mL_\tau
\end{equation}
and $\vec{e}:=(1,\dotsc,1)\in\mathbb{Z}^D$. With this choice, tensors of layers $\tau$ and $\tau'$ can end up at the same site only if
\begin{gather*}
\vec{r}_{\tau}(\vec{n}) = \vec{r}_{\tau'}(\vec{n}')\\
\Leftrightarrow\quad  n_i=b^{\tau'-\tau}n_i' + b^{\tau'-\tau-1}-b^{-1} \quad\forall_i,
\end{gather*}
 i.e., only if the two tensors belong to the same layer $\tau'=\tau$ and the same lattice cell $\vec{n}'=\vec{n}$ within that layer, following from the fact that the right hand side of the second equation is not an integer otherwise. The possible tensor positions of layers $\tau$ form disjoint sublattices $\V_\tau$ of the physical lattice $\V_\phys$.
\begin{gather}
	\V_\tau:=\{\vec{r}_{\tau}(\vec{n})|\vec{n}\in\mL_\tau\}\subset \V_\phys,\\
	\V_\tau\cap\V_{\tau'}=\emptyset\quad \forall_{\tau\neq\tau'}.
\end{gather}
All coordinates $r_i$ of $\vec{r}\in \V_\tau$ have a \emph{$b$-adic valuation} of $\tau-1$, where the $b$-adic valuation $v_b(n)$ of an integer $n$ is defined such that $v_b(n)=\tau$ iff $\tau$ is the largest integer such that $n\bmod b^{\tau}=0$, for example, $v_2(12)=2$.

\begin{figure}
\centering
\includegraphics[width=0.80\linewidth]{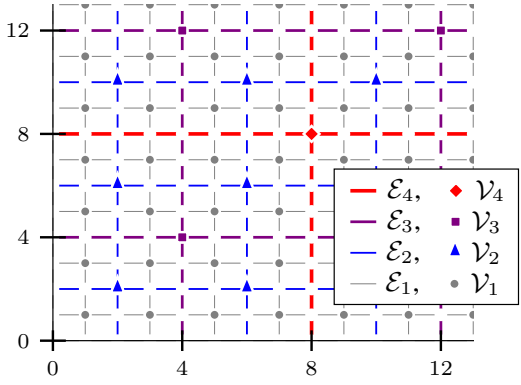}
\caption{\label{fig:dyadicGrid}
Disjoint sublattices $\V_\tau\subset\V_\phys$ to which MERA tensors are assigned and disjoint subsets of edges $\E_\tau\subset \E_\phys$ to which MERA contraction lines are assigned for $b=2$. In our construction, the paths assigned to contraction lines from tensors of layer 2 to tensors in layer 3 are for example restricted to edges from $\E_2\cup\E_3$.}
\end{figure}

Avoiding stacks of tensors is not sufficient for an efficient PEPS encoding. In $D=1$, all contraction lines are assigned to paths that necessarily stack up on the $x$-axis, Fig.~\ref{fig:MERA1d}. This stacking of the paths can be avoided in $D>1$ by assigning contraction lines between tensors of layers $\tau$ and $\tau'$ to paths that are restricted to edges from certain subgrids $\E_\tau$ and $\E_{\tau '}$ and that are shortest paths with respect to the $L_1$ distance on $\V_\tau\cup\V_{\tau '}$. Here, a grid $\E_\tau$ is defined as the subset of physical edges connecting nearest neighbors of the lattice $\V_\tau$ on straight lines; see Fig.~\ref{fig:dyadicGrid}.
\begin{gather}
	\E_\tau := \{(\vec{r},\vec{r}+\vec{e}_i)\in\E_\phys|\,v_b(r_j)=\tau-1\,\,\forall_{j\neq i}\}\\
	\Rightarrow\quad
	\E_\tau\cap \E_{\tau'} = \emptyset\,\, \forall_{\tau\neq \tau'}
\end{gather}
with $[\vec{e}_i]_j=\delta_{i,j}$.

For this choice for tensor positions and paths of MERA contraction lines an upper bound for the resulting PEPS refinement parameter $\chi_\peps$ follows: Contraction lines assigned to an edge $e=(\vec{r},\vec{r}')\in \E_\tau$ contract tensors of layer $\tau$ with tensors of layers $\tau'$ where $|\tau'-\tau|\leq C_T$. For a layer $\tau'$ with $\tau'>\tau$, tensors from at most $(2C_r)^D$ cells of $\mL_{\tau'}$ around the cell corresponding to site $\vec{r}$ can have contraction line paths traversing edge $e$. From the layers $\tau'$ with $\tau'\leq\tau$, tensors of at most $(2C_r)^D \sum_{t=0}^{C_T} b^{D t}$ cells can contribute. Thus, the number of contraction line paths traversing edge $e$ and hence $\log_\chi(\chi_\peps)$ are bounded from above by
\begin{equation}\label{eq:chiBound}
	\log_\chi(\chi_\peps) \leq (2C_r)^D (C_T+b^{D (C_T+1)})C_t C_o.
\end{equation}
As this upper bound is independent of the system size, the presented mapping of MERA to PEPS is efficient.

\subsection{Lower bond dimension by PEPS refinement}
\begin{figure}
\centering
\includegraphics[width=0.85\linewidth]{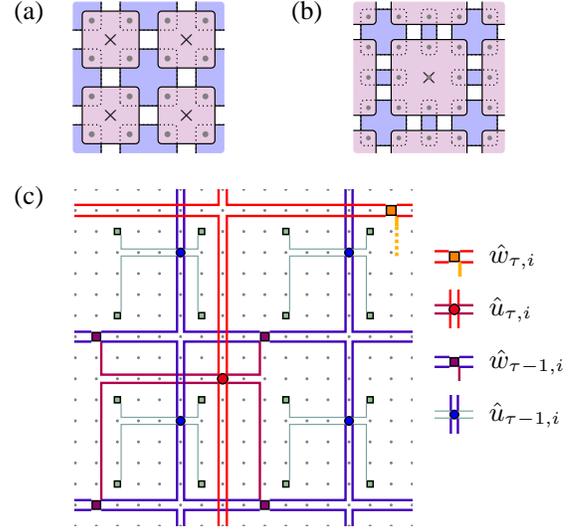}
\caption{\label{fig:MERA2d}
(a)~Unit cell of a specific 2D MERA state with linear branching ration $b=2$. With each layer, corresponding to a single renormalization step, unitary \emph{disentanglers} are applied that reduce the entanglement between blocks of $2\times 2$ sites with the rest of the system. Then, an isometry maps from those $2\times 2$ sites (dots) into one (crosses).
(b)~Unit cell of an alternative 2D MERA state. In each layer, unitary \emph{disentanglers} are applied that reduce the entanglement between blocks of $3\times 3$ sites with the rest of the system. Then, an isometry maps from those $3\times 3$ sites into one.
(c)~Mapping of the MERA state from (a) to a refined PEPS with $\delta \tau=1$. The diagram shows the assignment of two layers of the MERA, composed of disentanglers $\hat u$ and isometries $\hat w$, to the physical lattice. Each edge of the lattice is traversed by at most two contraction line paths. The resulting PEPS has $\chi_\peps=\chi^2$.}
\end{figure}
As it stands, for each layer, all tensors of a given lattice cell $\vec{n}\in\mL_\tau$ of layer $\tau$ are assigned to the same physical lattice site $\vec{r}_\tau(\vec{n})\in \V_\phys$ according to Eq.~\eqref{eq:tensorPos}. Therefore, a considerable number of contraction lines that start at the tensors of a given cell $\vec{n}$ may traverse the same edges around $\vec{r}_\tau(\vec{n})$ and cause hence a relatively high $\chi_\peps$. While this is unproblematic for the purpose of proving the existence of an efficient mapping, the situation can be improved for numerical purposes, e.g., by introducing for each site of the physical layer $b^{\delta\tau D}-1$ auxiliary sites with $\delta\tau>0$, resulting in \emph{refined} lattices $\V_\phys'$ and $\V_\tau'$. The corresponding \emph{refined PEPS} has tensors for the physical sites and tensors for the auxiliary sites, where the latter ones carry no physical indices. The sites from $\V_\phys'$ allowed for tensors of layer $\tau$ are then defined as
\begin{gather}\label{eq:tensorPosGen}
	\vec{r}_{\tau}(\vec{n},\vec{m}) := b^{\tau+\delta\tau}\vec{n}+b^{\tau}\vec{m}+b^{\tau-1}\vec{e}\quad \in \V_\phys'\\\nonumber
	 \quad\text{with}\quad \vec{n}\in \mL_\tau \quad\text{and}\quad m_i\in\{0,\dotsc,b^{\delta\tau}-1\}^D.
\end{gather}
Lattice cells in layer $\tau$ are again labeled by $\vec{n}\in\mL_{\tau}$ and $\vec{m}$ labels now the possible positions for tensors inside that cell.

Fig.~\ref{fig:MERA2d}c displays a mapping of the MERA from Fig.~\ref{fig:MERA2d}a to a refined PEPS with $\delta\tau=1$. Each edge of the grid is traversed by at most two contraction line paths. The resulting PEPS has consequently $\chi^\refined_\peps=\chi^2$. Due to the refinement of the physical lattice, the PEPS consists however of $|\V_\phys'|=b^{\delta\tau D}|\V_\phys|$ instead of $|\V_\phys|$ tensors.
A refined PEPS is transformed to a ``normal'' PEPS by contracting the PEPS tensors for the $b^{\delta\tau D}-1$ auxiliary sites with the tensor for the corresponding physical site, resulting in $\chi_\peps=(\chi^\refined_\peps)^{b^{\delta\tau(D-1)}}$.

\section{Examples for particular 2D MERA states}
The scheme displayed in Fig.~\ref{fig:MERA2d}c for mapping the 2D MERA defined in Fig.~\ref{fig:MERA2d}a to a PEPS results in the PEPS refinement parameter $\chi^\refined_\peps=\chi^2$ if one uses a refined PEPS with $\delta \tau=1$ where isometries are located at $\vec{m}=(1,1)$ and disentanglers at $\vec{m}=(0,0)$, according to Eq.~\eqref{eq:tensorPosGen}.
Using, instead, the tensor coordinates according to the most simple scheme, Eq.~\eqref{eq:tensorPos}, yields $\chi_\peps=\chi^6$.

An analogous scheme for mapping the slightly more complicated MERA defined in Fig.~\ref{fig:MERA2d}b to a PEPS results in the PEPS refinement parameter $\chi^\refined_\peps=\chi^5$ if one uses a refined PEPS with $\delta \tau=1$ where the isometries are located at $\vec{m}=(1,1)$, the $2\times 2$ disentanglers at $\vec{m}=(0,0)$, the $2\times 1$ disentanglers at $\vec{m}=(1,0)$, and the $1\times 2$ disentanglers at $\vec{m}=(0,1)$, according to Eq.~\eqref{eq:tensorPosGen}.

\section{1D TTN and MERA states cannot be mapped efficiently to 1D PEPS}\label{sec:1DcounterExample}
\begin{figure}
\centering
\includegraphics[width=1\linewidth]{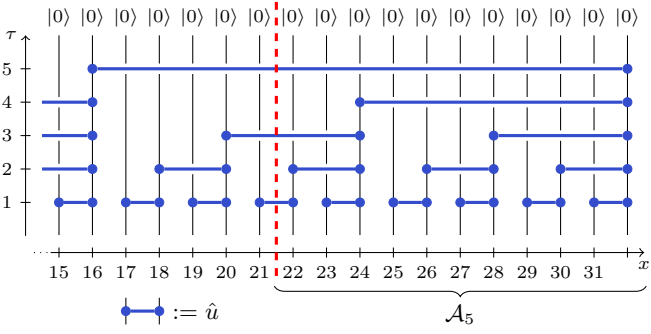}
\caption{\label{fig:TTNexample}
The 1D TTN state discussed in Sec.~\ref{sec:1DcounterExample} for $T=\log L=5$. For the chosen subsystems $\A_T$, the entanglement entropy is logarithmic in the system size, $S_{\A_T}=(T+1)/2$.}
\end{figure}
To show that a 1D MERA can in general not be mapped to a 1D PEPS with a bond dimension $\chi_\peps$ that is independent of the system size $L$, we construct a family of \emph{graph} TTN states \cite{Hein2004-69,Shi2006-74} for which the entanglement entropy grows logarithmically with $L$ for a suitable bipartition of the system; see, e.g., Refs.~\cite{Vidal2006,Evenbly2007,Dawson2008-100} for a numerical analysis.
We choose a TTN state, so a MERA state without disentanglers, with isometries mapping from two qubits to one, i.e., $\chi=2$ and $b=2$. Each representative of the family of states is defined on $L=2^T$ sites $\{0,\dotsc, L-1\}$ with a positive odd integer $T$. For simplicity, the TTN is embedded into the entire lattice of $L$ sites, and each isometry is considered as a unitary having one input from the previous layer and one input $|0\rangle$; see Fig. \ref{fig:TTNexample}. The state of the the top layer $\tau=T$ and the two-site gates $\hat u$ are defined as
\begin{equation}\label{eq:MERA1d-example}
	|\psi_{T}\ket:= | 0\ket^{\otimes L} \quad
	\text{and}\quad
	\hat u := e^{-i\pi \hat X\otimes \hat X/4},
\end{equation}
$\hat X$ and $\hat Z$ denoting Pauli matrices, and $|0\rangle$ being an eigenstate of $\hat Z$. For layers $\tau=1,\dots, T$ the state $|\psi_{\tau-1}\ket$ is generated from $|\psi_\tau\ket$ by applying gates $\hat u$ to sites $2^\tau(k-1/2)-1$ and $2^\tau k-1$ for $k=1,\dots, 2^{T-\tau}$. The entanglement is computed for the subsystem $\A_T$ consisting of the last $p_T$ sites, where $p_1:=1$ and $p_{T+2}:= 4p_{T}-1$. Since all the gates, specified in Eq.~\eqref{eq:MERA1d-example}, are mutually commuting, all gates that are supported entirely on $\A_T$ or entirely on its complement $\A_T^\bot$ can be disregarded for the computation of the entanglement entropy
\begin{equation*}
	S_{\A_T} = - \Tr \hat \rho_{\A_T} \log \hat \rho_{\A_T}
	\text{{ }with{ }}
	\hat \rho_{\A_T} = \Tr_{\A_T^\bot}|\psi_T\ket\bra \psi_T|.
\end{equation*}
The locations $p_T$ of the bipartitions are chosen such that, for odd $T$, exactly $(T+1)/2$ gates act across the cut. Since each such gate generates one pair of maximally entangled qubits, one obtains $S_{\A_T}= (T+1)/2$ which is logarithmically divergent in the system size $L$ and implies that any 1D PEPS encoding of the given graph state requires a $\chi_\peps$ that diverges with $L$.

\section{Efficiently contractible TNS that violate the area law}\label{sec:QCA-Graph}
As shown here, unlike for $D=1$ spatial dimensions, MERA states for $D>1$ obey the entanglement area law and not a log-area law, $S_{\A_L}= \Omega( L^{D-1}\log L )$, as it occurs for critical fermionic models \cite{Wolf2005,Gioev2005,Barthel2006-74,Li2006,Cramer2006,Amico2008-80,Eisert2008}. This raises the question of whether any efficiently contractible tensor network automatically yields an area law. This is however not the case.
In order to show this, we construct, for a $D$-dimensional cubic lattice of $L^D$ sites, a family of efficiently contractible TNS based on a unitary \emph{quantum cellular automata} (QCA). For a specific choice of the constituting tensors, one obtains instances that violate the area law for generic bipartitions of the system.

Let us consider a QCA consisting of $T$ layers $\tau=1,\dotsc ,T$, where each layer consists of two sublayers. With the first sublayer, $\hat K_1$ is applied which consists of $(L/2)^D$ local unitary gates $\hat s$ supported on plaquettes of $2\times\dotsb\times 2$ sites each. The operator $\hat K_2$ for the second sublayer is identical to $\hat K_1$ except for a relative shifting of the gate positions by $(1,\dotsc,1)$. Therefore, periodic boundary conditions are imposed, and $L$ is required to be even.
The initial state $|\psi_0\ket$ is a product state of $(L/2)^D$ plaquette states $|\phi\ket$ for $2\times\dotsb\times 2$ sites each, where the plaquette positions coincide with those of the gates in $\hat K_2$. The plaquette states are product states of $2^{D-1}$ maximally entangled pairs of qubits sitting each at the ends of the plaquette diagonals, e.g., $|\phi\ket=\hat u_{0,1}|0\ket^{\otimes 2}$ for $D=1$, where $\hat u$ is chosen according to Eq.~\eqref{eq:MERA1d-example} and the indices label the sites the gate acts on. For $D=2$, the plaquette state is $|\phi\ket=\hat u_{(0,0),(1,1)} \hat u_{(1,0),(0,1)}|0\ket^{\otimes 4}$.
The plaquette operators $\hat s$, composing the $\hat K_i$, are chosen as products of swap operators $\hat S_{i,j}|\sigma_i\sigma_j\ket=|\sigma_j\sigma_i\ket$ that act similarly on the qubits at the ends of the plaquette diagonals. For example, $\hat s = \hat S_{(0,0),(1,1)}\hat S_{(1,0),(0,1)}$ for $D=2$.

\begin{figure}
\centering
\includegraphics[width=1\linewidth]{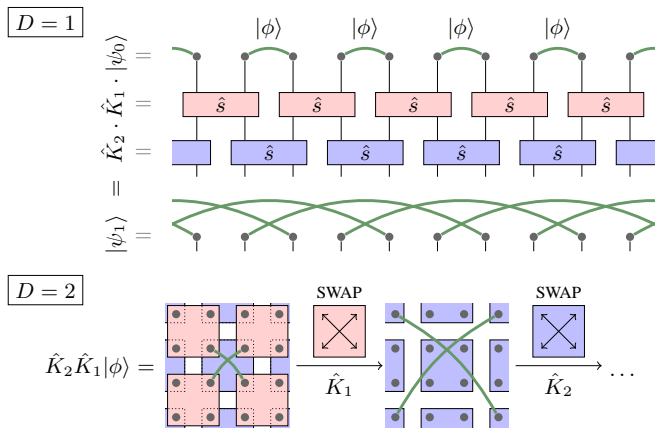}
\caption{\label{fig:QCA-Graph}
The graph state QCA discussed in Sec.~\ref{sec:QCA-Graph} for $D=1$ and $D=2$. Each layer of the QCA moves the maximally entangled qubits that initially are located on next nearest neighbor sites two steps further apart from each other. This increases the entanglement entropy for a given bipartition of the system in every step.
For an appropriate choice of the number of layers, the states violate the entanglement area law while still being efficiently contractible.}
\end{figure}

The QCA layers and the initial state are invariant under translations by two sites and rotations by $\pi/4$ and so are all states $|\psi_\tau\ket:=(\hat K_2\hat K_1)^\tau|\psi_0\ket$. Like $|\psi_0\ket$, every state $|\psi_\tau\ket$ is a product state of $2^{D-1}(L/2)^D$ maximally entangled qubit pairs. If two entangled qubits have positions $\vec{r}\pm \Delta\vec{r}$ in $|\psi_\tau\ket$ there is exactly one corresponding entangled qubit pair at positions $\vec{r}\pm (1+\frac{2\sqrt{D}}{|\Delta\vec{r}|})\Delta\vec{r}$ in $|\psi_{\tau+1}\ket$. Applying one QCA layer after another, distances of entangled qubits increase by $4\sqrt{D}$ in each step, e.g., $\hat K_1\hat K_2 \hat u_{(2,2),(3,3)} = \hat u_{(0,0),(5,5)}$, see Fig.~\ref{fig:QCA-Graph}.

For generic choices for bipartitions of the system into two parts, $\A_L\subset\V_\phys$ and its complement, where $\A_L$ is connected and has a volume $\propto L^D$, the corresponding entanglement entropy will violate the area law if the number of layers, $T$, is chosen appropriately. Consider as an example the bipartition with $\A_L=\{0,\dots,L/2-1\}\times\{0,\dots, L-1\}^{D-1}$. The subsystem boundary is formed by the planes $\{\vec r|r_1=0\}$ and $\{\vec r|r_1=L/2\}$. Each plane is crossed by a number of different pairs of entangled qubits that is proportional to its area and to $T$, as long as $L/2>2T$. Consequently,
\begin{equation}\label{eq:qcaEntropy}
	S_{\A_L}(T)=\Omega(L^{D-1} T).
\end{equation}

For a choice $T\propto \log L$, this yields a log-area law $S_{\A_L}=\Omega(L^{D-1} \log L)$. But an upper bound on the computation cost for the evaluation of local observables with respect to QCA states of the given class (with arbitrary $\hat u$) is of order $O(2^{2 T^D} T)$, i.e., $O(L^{2(\log L)^{D-1}}\log L)$ for $T\propto\log L$. This cost is not polynomial in $L$ and the QCA are for this $T$ hence not efficiently contractible in an obvious fashion.
However, the computation cost is, of order $O(L^2 (\log L)^{1/D})$ for the choice $T=(\log L)^{1/D}$, i.e., polynomial in $L$. The resulting entanglement entropy is $S_{\A_L}=\Omega(L^{D-1}(\log L)^{1/D})$ which violates the area law by the sublogarithmic factor $(\log L)^{1/D}$.
Note also that even a QCA consisting of a single layer of $k\times \dots\times k$ plaquettes, where $k$ is allowed to grow as $k\propto (\log L)^{1/D}$, can also violate the entanglement area law, albeit only for specific bipartitions of the system, while being efficiently contractible.

\vspace{0.5cm}
\section{Conclusion}
In this work, we have shown that MERA states for $D>1$ can be encoded efficiently as PEPS. From the perspective of numerical simulations for strongly correlated many-body systems, this means that $D>1$ MERA states form a subclass of efficiently contractible PEPS. From a physical perspective, the result implies that real-space renormalization techniques, despite the scale-invariant features of the TNS they generate, give rise to states that can be encoded with \emph{local} degrees of freedom. As a corollary, it follows that $D>1$ MERA states feature an \emph{area law} for the entanglement entropy \cite{Amico2008-80,Eisert2008,Vidal2006arXiv}. Consequently, the refinement parameter $\chi$ needs to be scaled polynomially in the system size in order to describe critical fermionic systems accurately. Otherwise, the real-space RG schemes addressed here \cite{Jullien1977-38,Drell1977-16,Vidal-2005-12} are imprecise for such systems.
Constructing explicit examples, it has been shown further that 1D MERA states can in general not be encoded efficiently as PEPS and that, for $D>1$, there exist efficiently contractible TNS, based on quantum cellular automata, that violate the entanglement area law.
It is the hope that this work further clarifies the intricate structure of the small number of effective degrees of freedom that typical ground-states of local quantum many-body systems explore.

\acknowledgments
We thank V.\ Nesme, A.\ Flesch, and G.\ Vidal for fruitful discussions.
This work has been supported by the EU (MINOS, QESSENCE, COMPAS), and the EURYI.

\clearpage
\bibliographystyle{prsty} 

\end{document}